\documentclass[preprint]{aastex}
\usepackage{natbib}
\shorttitle{Acceleration of Type 2 Spicules-2}
\shortauthors{M.L. Goodman} 

\begin{document}

\title{Acceleration of Type 2 Spicules in the Solar Chromosphere - 2: Viscous Braking and Upper Bounds on Coronal Energy Input} 
\author{Michael L. Goodman}        
\affil{Advanced Technologies Group, West Virginia High Technology Consortium Foundation\\1000 Galliher Drive, Fairmont, WV 26554}
\email{mgoodman@wvhtf.org \\ ApJ - Accepted 2/26/2014}


\begin{abstract}
A magnetohydrodynamic model is used to determine conditions under which the Lorentz force accelerates plasma to type 2 spicule speeds in the chromosphere. The model generalizes a previous model to include a more realistic pre-spicule state, and the vertical viscous force. Two cases of acceleration under upper chromospheric conditions are considered. The magnetic field strength for these cases is $\leq 12.5$ and 25 G. Plasma is accelerated to terminal vertical speeds of 66 and 78 km-s$^{-1}$ in $100$ s, compared with 124 and 397 km-s$^{-1}$ for the case of zero viscosity. The flows are localized within horizontal diameters $\sim 80$ and 50 km. The total thermal energy generated by viscous dissipation is $\sim 10$ times larger than that due to Joule dissipation, but the magnitude of the total cooling due to rarefaction is $\gtrsim$ this energy. Compressive heating dominates during the early phase of acceleration. The maximum energy injected into the corona by type 2 spicules, defined as the energy flux in the upper chromosphere, may largely balance total coronal energy losses in quiet regions, possibly also in coronal holes, but not in active regions. It is proposed that magnetic flux emergence in intergranular regions drives type 2 spicules.
\end{abstract}

\section*{1. Introduction}

In a review of small scale chromospheric structures, Tsiropoula et al. (2012) point out that the existence of spicules in the chromosphere, eventually identified as mainly vertical supersonic flows, has been inferred from observations for more than 130 years. As spatial and temporal resolution has
increased, the observed maximum flow speeds of spicules have increased, and the observed minimum length scales over which these flows are
localized orthogonal to their velocity have decreased. This trend is similar to that of observing, and using semi-empirical models to infer the existence of, increasingly stronger magnetic field strengths and total magnetic energy on increasingly smaller horizontal scales in the photosphere (de Wijn et al. 2009; S\'{a}nchez Almeida \& Mart\'{i}nez Gonz\'{a}lez 2011; Stenflo 2012, 2013). These trends can be connected in that Lorentz forces on a given scale can drive flow on that scale. There is no sign that the smallest scales of magnetic fields or supersonic flows in any region of the solar atmosphere are resolved.

These trends suggest that type 2 spicules, observed on the limb and on the disk (as rapid blueshifted excursions - RBE's) (De Pontieu et al. 2007 a,b,c; Langangen et al. 2008; Rouppe van der Voort et al. 2009; Sekse et al. 2012; De Pontieu et al. 2012; de Wign 2012) are the sub-set of chromospheric flows with the highest speeds $(\sim 50-150 \; \mbox{km-s$^{-1}$})$, and shortest durations $(\sim 10-150 \; \mbox{s})$ that can be detected with the highest available spatial and temporal resolution of $\sim 100 \; \mbox{km}$ and 5-8 s (van Noort \& Rouppe van der Voort 2006; De Pontieu et al. 2007 c; Tavabi et al. 2011; Pereira et al. 2012).

This paper uses a magnetohydrodynamic (MHD) model that improves upon the model in Goodman (2012, henceforth G12) to determine conditions under which flows can be accelerated to vertical speeds comparable to type 2 spicule speeds in the chromosphere. The model in G12 assumes a pre-spicule state without flow, with a constant vertical magnetic field, and neglects the effects of viscosity. The model presented here includes a pre-spicule state with 2 D velocity and magnetic fields, and the viscous force due to HI-HI collisions. 

\section*{2. The Model}

Assume cylindrical coordinates $(R,\theta,z)$. Let $t$ be time. All quantities are independent of $\theta$, and $z$ is height above some to be specified point in the chromosphere. Assume a weakly ionized H plasma with constant temperature $T$. Then $p=\rho k_B T/m_p= \rho V_s^2$, where $p,\rho, V_s, m_p$, and $k_B$ are the pressure, mass density, sound speed, proton mass, and Boltzmann's constant. Assume $p=p_1(R,t) \exp(-z/L)$ and $\rho=\rho_1(R,t) \exp(-z/L)$. Here $L=k_B T/m_p g$ is the pressure scale height, where $g=2.74 \times 10^4$ cm-s$^{-2}$. Then the vertical pressure gradient and gravitational forces cancel.

Let ${\bf V}$ be the center of mass (bulk flow) velocity. Assume the magnetic field ${\bf B} = {\bf b}(R,t) \exp(-z/2L)$. Include the viscous force due to HI-HI collisions in the momentum equation. For the momentum and mass conservation equations an exact solution for the height dependence of ${\bf V}$
is that it is independent of height. This form of the solution is assumed here, so ${\bf V}={\bf V}(R,t)$. The assumed form of the height dependence of ${\bf B}, {\bf V}$, and $\rho$ implies the flow is accelerated by height independent Lorentz and viscous forces per unit mass acting on a vertical column of gas. The current density ${\bf J} = {\bf j}(R,t) \exp(-z/2L)  = c \nabla \times {\bf B}/4 \pi$. It is assumed that $V_\theta=0$. 

Under the preceding assumptions, the momentum and mass conservation equations are
\begin{equation}
\dot{V}_R + V_R V_R^\prime + V_s^2 \frac{\rho_1^\prime}{\rho_1}= \frac{j_\theta b_z-j_z b_\theta}{c \rho_1} + 
\frac{4 \nu}{3 \rho_1} \left(V_R^{\prime \prime} + \frac{V_R^\prime}{R} - \frac{V_R}{R^2}\right). \label{1}
\end{equation}
\begin{equation}
j_z b_R - j_R b_z=0. \label{2}
\end{equation}
\begin{equation}
\dot{V}_z + V_R V_z^\prime= \frac{j_R b_\theta-j_\theta b_R}{c \rho_1} + \frac{\nu}{\rho_1 R} \left(R V_z^\prime\right)^\prime. \label{3}
\end{equation}
\begin{equation}
\dot{\rho}_1 + \rho_1 \frac{(R V_R)^\prime}{R} +V_R \rho_1^\prime = \frac{\rho_1 V_z}{L}. \label{4}
\end{equation}
Here the dot and prime denote $\partial/\partial t$ and $\partial/\partial R$, and $\nu=5(m_p k_B T)^{1/2}/(16 \pi^{1/2} d^2)$ is the viscosity of HI,
where $d$ is an estimate of the atomic diameter (Mihalas \& Mihalas 1984). Here $d=2 a_0$, where $a_0$ is the Bohr radius.

The $|V_R|$ is now restricted to being sufficiently small to simplify solving the model. This restriction, together with the restriction made above that $V_\theta =0$, is consistent with the observation that the primary acceleration is in the vertical direction. These restrictions prevent the
model from reproducing horizontal and torsional flows, with speeds up to $\sim 30$ km-s$^{-1}$, exhibited by some type 2 spicules (Sekse et al.
2012; De Pontieu et al. 2012).

Constraints are imposed on $V_R$ to justify omitting the terms involving $V_R$ in the momentum equations (\ref{1}) and 
(\ref{3}). Inspection of those equations indicates the constraints $|V_R| \ll |V_z|$, and $|V_R| < V_s$ should be adequate for the solutions presented in \S 6. As shown in \S 3, the constraint on $V_R$ places an upper bound on $R_0$, which is the characteristic radius within which ${\bf J}$ is confined. Imposing these constraints reduces equations (\ref{1}) and (\ref{3}) to
\begin{equation}
V_s^2 \rho_1^\prime= \frac{j_\theta b_z-j_z b_\theta}{c}, \label{5}
\end{equation}
\begin{equation}
\dot{V}_z = \frac{j_R b_\theta-j_\theta b_R}{c \rho_1} + \frac{\nu}{\rho_1 R} \left(R V_z^\prime\right)^\prime. \label{6}
\end{equation}
In order that the solution for $V_z$ be finite at $R=0$ it is necessary that the boundary condition $V_z^\prime(0,t)=0$ be imposed. One more boundary
condition is necessary to uniquely determine the solution for $V_z$. This is chosen as $V_z(R= 100 \; \mbox{km},t)=0$ since the observed diameter of type 2 spicules is $\lesssim 200$ km. For the solutions in \S 6, the plasma acceleration, and resistive, viscous, and compressive heating occur almost entirely within a radius $\ll 100$ km from the origin. 

\subsection*{2.1. Accelerating and Compressive Lorentz Forces}

If the Lorentz force plays an important role in accelerating spicules then it must be localized on the space and time
scales observed to characterize spicules. The vertical Lorentz force is $(J_R B_\theta - J_\theta B_R)/c$. Here a functional form for $j_z$ is assumed, given by Eq. (\ref{7}) below, that is localized on time and radial scales $t_0$ and $R_0$ to model the temporal and radial localization of type 2 spicules. As shown below, the assumed form for $j_z$ together with the equations $\nabla \cdot {\bf J}=0$, and $\nabla \times {\bf B}=4 \pi {\bf J}/c$ determine $j_R$ and $b_\theta$, and hence the contribution $J_R B_\theta/c$ to the vertical Lorentz force. 
For the two spicule solutions presented in \S 6, this is the accelerating Lorentz force in that it drives all of the upward acceleration of the plasma. It does this against a much weaker downward component, $-J_\theta B_R/c$, of the vertical Lorentz force present in the
background (BG) state. The values of $t_0$ and $R_0$ are chosen so the solutions exhibit a plasma acceleration time, and radial localization of the accelerated plasma consistent with observations. The observations of type 2 spicules referenced in \S 1 show they have durations $\sim 10 - 150$ s, and characteristic diameters $\stackrel{<}{\sim} 200$ km. In \S 6, the choice $t_0=33.3$ s yields an acceleration time of $\sim 100$ s to maximum vertical speeds of $\sim 66$ and 77 km-s$^{-1}$, and the corresponding choices of $R_0 = 5$ and 1.25 km yield flows mainly confined within diameters 
$\sim 80$ and 50 km.

$j_z$ is assumed to have the following form.
\begin{equation}
j_z = j_0 \exp(-\bar{t})(1-\exp(-\bar{t})) \exp(-\bar{R}^2). \label{7}
\end{equation} 
Here $\bar{R} = R/R_0$ and $\bar{t} = t/t_0$. 

The specification of $J_z, T$, and $V_\theta$ causes the complete MHD model to be over-determined. Consequently, three equations must be omitted from the complete MHD model. The remaining equations determine the solution self-consistently. Here the mass and momentum equations, Faraday's law, the $R$ component of a multi-fluid Ohm's law, the ideal gas equation of state, and an NLTE Saha equation are used to determine the solution to the model. The energy equation, and the $\theta$ and $z$ components of the Ohm's are omitted from the model,
except that the complete Ohm's law is used to estimate the Joule heating rate.

It follows from $\nabla \cdot {\bf J}=0$ that 
\begin{equation}
j_r = \frac{j_0 R_0}{4 L} \exp(-\bar{t})(1-\exp(-\bar{t})) \frac{(1-\exp(-\bar{R}^2))}{\bar{R}}. \label{8}
\end{equation} 

From $\nabla \times {\bf B} = 4 \pi {\bf J}/c$ it follows that
\begin{equation}
b_\theta=b_{\theta 0} \exp(-\bar{t})(1-\exp(-\bar{t})) \frac{(1-\exp(-\bar{R}^2))}{\bar {R}}, \label{9}
\end{equation} 
where $b_{\theta 0} \equiv 2 \pi j_0 R_0/c$. 

With respect to $\bar{t}$, the maximum values of $j_z,j_r$, and $b_\theta$ are reached at $\bar{t}=\ln(2) \sim 0.693147$, for which 
$\exp(-\bar{t})(1-\exp(-\bar{t})) =  1/4$. With respect to $\bar{R}$, the maximum values of $j_r$ and $b_\theta$ are reached at $\bar{R}=  1.1209$, for
which $(1-\exp(-\bar{R}^2))/\bar{R} = 0.6382$. The maximum value of $b_\theta$, and hence of $B_\theta$ is then $0.6382 \; b_{\theta 0}/4 = 0.1596 \; 
b_{\theta 0}$. Then specifying $b_{\theta 0}$ determines the maximum magnitude of $B_\theta$, and determines $j_0 R_0$. Figure 1 shows 
$B_\theta/B_{\theta, max}$ over the effective time interval $0 \leq \bar{t} \leq 4$ of Lorentz force driven acceleration,
where $B_{\theta, max}$ is the maximum value of $B_\theta$.

\subsection*{2.2. Background (BG) State}

This is a state in which $B_\theta=0$. It is determined by $B_R,B_z$, and $J_\theta$ in that once these quantities are known they determine $\rho, V_z$,
and $V_R$ for this state. Equation (\ref{2}), and $\nabla \cdot {\bf B}=0$ imply
\begin{eqnarray}
B_R(R,z,t)&=& \frac{B_0(t) R_0}{4 L \bar{R}} \left(1-\exp(-\bar{R}^2)\right) \exp(-z/2L) \label{10} \\
B_z(R,z,t) &=& B_0(t) \exp(-\bar{R}^2) \exp(-z/2L) \label{11},
\end{eqnarray}
where $B_0(t)$ is an arbitrary function of time, specified in \S 3. For the spicule solutions presented in \S 6, $B_\theta \gtrsim 10 (B_R^2 + 
B_z^2)^{1/2}$ during the main phase of the acceleration. Such a highly twisted field may be susceptible to the kink instability in the real chromosphere. The specification of the time dependence of $j_z$ and $B_0$ forces the model solutions to be stable. A model driven only by initial and boundary conditions is required to determine the stability of such field configurations in the chromosphere.

From equations (\ref{10}) and (\ref{11}) it follows that
\begin{equation}
J_\theta = \frac{- B_0(t) c \bar{R}}{2 \pi R_0} \left[\left(\frac{R_0}{4 L \bar{R}}\right)^2 \left(1-\exp(-\bar{R}^2)\right) - \exp(-\bar{R}^2)\right]
\exp(-z/2L) \label{12}
\end{equation}

The simplifying assumptions in \S 2 that $V_\theta=0$ and $\partial/\partial \theta =0$ cause the $\theta$ component of the momentum equation to reduce to Eq. (\ref{2}). This causes the BG state to become un-coupled from the spicule in that the BG state influences the spicule acceleration process, but this process does not affect the BG state. In reality, there is bi-directional coupling between the pre-spicule state of the atmosphere, and the subsequent spicule acceleration process. The strength of this coupling is not known. A model that allows for bi-directional coupling, for example one that does not require that $V_\theta=0$, or that $V_R$ and $R_0$ be small in the sense defined in \S\S 2 and 3, is needed to estimate the importance of this coupling. A potentially important effect omitted due to the un-coupling of the spicule from the BG plasma is the distortion of the BG magnetic field due to partial flux freezing, which increases with temperature and degree of ionization.  

The model presented here assumes the existence of a current that accelerates plasma through the Lorentz force. The source of this current is not specified. It is proposed here that the source is the emergence of current carrying magnetic flux through the photosphere in inter-granular 
lanes on scales $\lesssim 10^2$ km. These magnetic structures may rise into the chromosphere with unbalanced Lorentz forces that accelerate plasma as part of the process of relaxation of ${\bf B}$ towards a force free, minimum energy state. Magnetic flux emergence, driven by convection, occurs
continuously on inter-granular scales, and as observations have improved, more flux has been observed on increasingly smaller scales (Lites et al. 1996; Orozco Su\'{a}rez et al. 2007; Lites et al. 2008; Lites 2009). These emerging magnetic structures have or develop a vertical, quasi-cylindrical geometry, similar to the geometry of type 2 spicules. The proposition that magnetic flux emergence drives type 2 spicules is discussed in more detail in 
\S 7. 

Driving by magnetic reconnection (e.g. Rouppe van der Voort et al. 2009) is unlikely since type 2 spicules are often observed in regions that appear to be unipolar (McIntosh \& De Pontieu 2009; Mart\'{i}nez-Sykora et al. 2011), and since in reconnection an upward jet is expected to be accompanied by a downward jet, whereas there do not appear to be observations of type 2 spicules as multi-jet phenomena. The model presented here includes a radial Lorentz force that compresses the plasma, increasing its density by factors up to $\sim 10^2$, but the model excludes the possibility
of this compression causing a vertical pressure gradient that contributes to spicule acceleration. It is possible that such a pressure gradient is generated in a more realistic model. MHD simulations support this possibility, showing the development of type 2 spicule like jets in regions with intense electric currents associated with magnetic flux emergence on granulation spatial scales (Mart\'{i}nez-Sykora et al. 2009; Mart\'{i}nez-Sykora et al. 2011).

\section*{3. General Solution for $\rho$ and ${\bf V}$}

Solving Eq. (\ref{5}) for $\rho_1$ gives
\begin{equation}
\rho_1 = \rho_{1 \infty} \left[1+  A(\bar{R})((b_{\theta 0}\exp(-\bar{t})(1-\exp(-\bar{t})))^2 + (\alpha B_0)^2) - \frac{B_0(t)^2}{8 \pi V_s^2 
\rho_{1 \infty}} \exp(-2 \bar{R}^2) \right]. \label{13}
\end{equation}
Here $\alpha = R_0/4L$,
\begin{eqnarray}
A(\bar{R})&=&\frac{I_0(\bar{R})}{4 \pi V_s^2 \rho_{1 \infty}}, \label{14} \\
I_0(\bar{R})&=& \int_{\bar{R}^2}^\infty \frac{(1-\exp(-y))\exp(-y)}{y} \; dy, \label{15}
\end{eqnarray}
and $\rho_{1 \infty} \equiv \rho_1(R=\infty)$ is the time independent BG density far from the acceleration region.

It is assumed the BG state evolves on the characteristic granulation turnover timescale $t_{bg} = 600$ s. Then this state evolves slowly compared with the spicule acceleration timescale of $\sim 100$ s for the solutions in \S 6. Choose $B_0(t)= b_0 \exp(-t/t_{bg}) = b_0 \exp(-\bar{t} 
t_0/t_{bg})$, where $b_0$ is constant. The choice of $b_0$ is constrained by the requirement that $\rho > 0$, and is discussed in \S 4. 

With $\rho_1$ known, $V_z$ is determined by numerically integrating Eq. (\ref{6}). This is done using the Matlab function pdefun that is adaptive in time, and uses a fixed, specified spatial grid. Here the spatial grid is specified by the range $0 \leq R \leq 100$ km, and grid spacing $0.005 R_0$. From Eq. (\ref{4}), 
\begin{equation}
V_R=\frac{R_0}{\rho_1(\bar{R},t) \bar{R}} \int_0^{\bar{R}} x \left(\frac{\rho_1(x,t) V_z(x,t)}{L} - \dot{\rho}_1(x,t) \right) \; dx. \label{19}
\end{equation}
It is found numerically that $V_R$ increases with $R_0$, roughly linearly for the solutions presented in \S 6.\footnote{This behavior can be shown analytically when $\nu=0$.} This is where the constraint on $R_0$ enters the solution. For given values of the other input parameters, $R_0$ must be chosen sufficiently small so that $|V_R| \ll |V_z|$, and $|V_R| < V_s$. 

\section*{4. Ohm's Law and Joule, Compressive, and Viscous Heating Rates}

The electron and HI number densities $n_e$ and $n_H$ are needed to compute terms in the Ohm's law. Here $n_e$ is computed using the NLTE, statistical equilibrium Saha equation derived in Goodman \& Judge (2012, Eq. 11), and $n_H$ is estimated as $\rho/m_p$.

The Ohm's law for the partially ionized plasma is assumed to be (Mitchner \& Kruger 1973)
\begin{equation}
{\bf E} + \frac{ {\bf V} \times {\bf B}}{c} = \eta \left({\bf J} + M_e \frac{{\bf J} \times {\bf B}}{B} + \Gamma {\bf J}_\perp\right). \label{300}
\end{equation}
Here $\eta$ is the Spitzer resistivity, ${\bf J}_\perp$ is the component of ${\bf J} \perp {\bf B}$, the magnetization factor $\Gamma = M_e M_p 
(\rho_n/\rho)^2$, where $M_e, M_p$, and $\rho_n$ are the electron and proton magnetizations\footnote{The magnetization of a particle species is the ratio of its cyclotron frequency to its total momentum transfer collision frequency.}, and the neutral density. In the chromosphere $\rho_n \sim \rho$. Explicitly,
\begin{eqnarray}
\eta &=& m_e^{1/2} \left(\frac{4 (2 \pi)^{1/2} e^2 \ln(\lambda)}{3 (k_B T)^{3/2}} + \frac{\sigma \rho (k_B T)^{1/2}}{n_e m_p e^2}\right), \label{25} \\
\lambda &=& \frac{3(k_B T)^{3/2}}{2 e^3(\pi n_e)^{1/2}},  \label{26} \\
\eta \Gamma &=& \frac{B^2 m_p^{1/2}}{c^2 \sigma \rho n_e (k_B T)^{1/2}}, \; \mbox{and}  \label{27} \\
\frac{\eta M_e}{B} &=& \frac{1}{n_e e c}. \label{200}
\end{eqnarray}
Here $m_e, e$, and $\sigma (= 5 \times 10^{-15}$ cm$^2$) are the electron mass and charge magnitude, and the charged-neutral particle scattering cross 
section (Osterbrock 1961). The MHD Joule heating rate $Q_J = {\bf J} \cdot ({\bf E} + ({\bf V} \times {\bf B})/c)$, where ${\bf E}$ is the electric field.
From the Ohm's law, $Q_J = \eta (J^2 + \Gamma J_\perp^2)$. The Pedersen current dissipation rate $Q_P \equiv \eta \Gamma J_\perp^2$, where $\eta \Gamma$ is the magnetization induced resistivity. 
The second term on the right hand side of the Ohm's law Eq. (\ref{300}) is the Hall electric field. It is not dissipative since it is $\perp {\bf J}$. It is used in the model to compute $E_R$, which is used in
\S 6.1.3 to compute the $z$ component of the Poynting flux.

The compressive heating rate per unit volume $Q_{comp} = - p \nabla \cdot {\bf V}$.

The viscous heating rate per unit volume may be written as\footnote{Exact expressions for the viscous heating rate and the viscous stress tensor in cylindrical and spherical coordinates in 3D are given in Appendix E of Thompson (1988). Here the continuity equation and the expression for $Q_{comp}$ are used to re-write the viscous heating rate in the form of Eq. (\ref{1002}) to simplify computation.}
\begin{equation}
Q_{vis} = \nu \left( {V_z^\prime}^2 + \frac{4}{3} \left(\frac{Q_{comp}}{\rho_1 V_s^2}\right)^2 + \frac{4 V_R}{R} \left(\frac{V_R}{R} - 
\frac{V_z}{L} + \frac{\left(\dot{\rho_1}   + \rho_1^\prime  V_R \right)}{\rho_1} \right) \right)  \label{1002}
\end{equation}

\section*{5. Faraday's Law}

Faraday's law reduces to 
\begin{eqnarray}
E_\theta(R,t) &=& - \frac{2 L}{c} \dot{B_R}, \label{100} \\
E_z(R,t) &=& E_z(0,t) + \int_0^R \left(\frac{\dot{B_\theta}}{c} - \frac{E_R}{2 L} \right) d\alpha, \label{101}
\end{eqnarray}
where $E_z(0,t)= \eta(0,t) J_z(0,t)$, and $E_R$ is assumed to be given by the $R$ component of the Ohm's law Eq. (\ref{300}).

\section*{6. Particular Solutions - Upper Chromosphere}

The solutions in this section are for spicules accelerated in the upper chromosphere.
The inputs to the model are $R_0, t_0, t_{bg}, b_{\theta 0}, b_0, \rho_{1 \infty},L$, and $T$. Two solutions are presented. For both solutions,
$t_0=33.3$ s, $t_{bg}=600$ s, and $T=8000$ K. Then $L = 241$ km, $V_s =8.13$ km-s$^{-1}$, and $\nu = 2.1 \times 10^{-3}$ poise.

The choice of $b_0$ is constrained by the requirement $\rho_1(R,t) > 0$. Equation (\ref{13}) shows that $\rho_1$ is a minimum in the BG state,
for which $b_{\theta 0}=0$. For the solutions considered here, $R_0 = 1.25$ km or 5 km. This determines the possible values of $\alpha(=R_0/4L)$. It can then be shown from Eq.(\ref{13}) with $b_{\theta 0}=0$ that for these values of $\alpha$, $d \rho_1/d\bar{R} > 0$ for 
$0 \leq \bar{R}< \bar{R}_\ast$, where $\bar{R}_\ast \sim 2.92-3.82$, and that $d \rho_1/d\bar{R} \leq 0$ for $\bar{R} \geq \bar{R}_\ast$, where equality holds only at $\bar{R} = \bar{R}_\ast$. However, for $\bar{R} = \bar{R}_\ast$, $\rho_1$ is already essentially equal to $\rho_{1 \infty}$. It follows that
the minimum value of the BG density occurs at $\bar{R}=0$. Then, noting that the maximum value of $B_0$ is $b_0$,
a necessary and sufficient condition for $\rho_1$ to be positive is
\begin{equation}
b_0 < 2 V_s \left(\frac{2 \pi \rho_{1 \infty}}{1 - 2 \alpha^2 I_0(0)}\right)^{1/2}, \label{23}
\end{equation}
where $I_0(0)=0.6931$. Since $\alpha^2 \sim 10^{-5}- 10^{-3}$ for the values of $R_0$ and $L$ used here, the denominator in 
Eq. (\ref{23}) is essentially unity. Then Eq. (\ref{23}) reduces to $b_0 < 2 V_s \left(2 \pi \rho_{1 \infty}\right)^{1/2} \sim 1.6662 
(n_{\infty}/10^{11} \mbox{cm$^{-3}$})^{1/2}$ G, where $n_\infty = \rho_{1 \infty}/m_p$. For upper chromospheric densities this constrains $b_0$ to be no more than a few Gauss. 

For all solutions $n_\infty=2.575 \times 10^{11}$ cm$^{-3}$. The constraint on $b_0$ is then $b_0 < 2.674$ G. Choose $b_0 = 2.5$ G for all solutions.
This implies the density at $\bar{R}=\bar{t}=0$ is $3.24 \times 10^{10}$ cm$^{-3}$. 

It remains to specify $b_{\theta 0}$ and $R_0$ to determine the solutions. All plots are at the reference height $z=0$. The reference height is the height at which $n(R,z,t)=n_\infty$ at $t=0$ and $R=\infty$. This follows from Eq. (\ref{13}) and $\rho(R,z,t)=\rho_1(R,t) \exp(-z/L)$.

The total magnetic field strengths for the spicule solutions presented here and in G12 during the main phase of the
acceleration are $\sim 10 - 25$ G. This is within the range inferred from observations of spicules, typically at heights 
$\sim 1400 - 5000$ km above the photosphere (L\'{o}pez Ariste \& Casini 2005; Trujillo Bueno 2005; Ramelli et al. 2006; Centeno et al. 2010). 

\subsection*{6.1. Solution 1}

For this solution $b_{\theta 0}=78.35$ G and $R_0=5$ km. Then $B_\theta \leq 12.5$ G. 

\subsubsection*{6.1.1. BG State}

This state is unremarkable for the given input parameters. The magnetic field is essentially vertical. $V_z$ and $V_R$ are $< 0$, with maximum magnitudes 2.2 km-s$^{-1}$ and 34 m-s$^{-1}$. There is a density depression localized in the region $\bar{R} \lesssim 0.7$, with the maximum depression at $R=0$. The density at $R=0$ is $\sim 3 \times 10^{10}$ cm$^{-3}$. It increases to $1.13 \times 10^{11}$ cm$^{-3}$ at $\bar{t}=4$, and continues to increase 
towards $n_\infty$ as the BG magnetic field decays.

The total thermal energy $E_J$ generated by Joule heating in the BG state is estimated as the integral of $Q_J$ over the volume $V$ defined
by $0 \leq z \leq \infty, 0 \leq R \leq 100$ km, and $0 \leq \theta \leq 2 \pi$, and over the time interval $T$ defined by $0 \leq t \leq 4 t_0$. Then $E_J = 9.14 \times 10^{18}$ ergs. Since $\bf{J} \perp {\bf B}$, $E_J$ is entirely due to Pedersen current dissipation.

The MHD Poynting theorem may be written as $(8 \pi)^{-1}\partial B^2/\partial t + \nabla \cdot {\bf S} = - Q_J - R_{ke}$. Here $R_{ke} \equiv
{\bf V} \cdot ({\bf J} \times {\bf B})/c$ is the rate per unit volume at which electromagnetic energy is transformed into bulk flow kinetic energy by the action of the Lorentz force, ${\bf S}$ is the Poynting flux, and $Q_J + R_{ke} = {\bf J} \cdot {\bf E}$. The total amount of electromagnetic energy transformed into bulk flow kinetic energy is denoted by $W$, and is defined as the integral of $R_{ke}$ over $V$ and $T$. Then $W= - 3.84 \times 10^{17}$ ergs. The negative value of $W$ indicates that bulk flow kinetic energy is converted into electromagnetic energy. The net amount of electromagnetic energy converted into particle energy within $V$ during $T$ is then $E_J+W \sim 8.76 \times 10^{18}$ ergs.

Similarly, the thermal energy generated by viscous and compressive heating is estimated as the integrals of $Q_{vis}$ and $Q_{comp}$ over $V$ and $T$, and denoted $E_{vis}$ and $E_{comp}$. For the BG state $(E_{vis}, E_{comp}) = (1.52 \times 10^{18}, 2.01 \times 10^{19})$ ergs.

\subsubsection*{6.1.2. Spicule Acceleration}

Set $b_{\theta 0}=78.35$ G. Then $B_\theta \leq 12.5$ G. Again generate the solution for $0 \leq \bar{t} \leq 4$.

Figure 2 shows the number density $n = \rho/m_p, V_z$, and $V_R$. During acceleration the plasma is compressed to a
density $\sim 100$ times its pre-acceleration value. The compression is confined within a diameter $\sim 20$ km, and is a maximum at $t \sim 0.6932 \; 
t_0 = 23$ s. The maximum of $V_z= 65.6$ km-s$^{-1}$. If the viscous force is neglected, the maximum value of $V_z$ is 124.1 km-s$^{-1}$.
The constraint $|V_R|\ll V_z$ is satisfied almost everywhere, in particular where almost all acceleration occurs and where $V_z$ is a maximum.
The maximum radial speed is 4.4 km-s$^{-1}$, and the radial flow changes from an inflow to an outflow during the acceleration. The flow is localized within a diameter $\sim 80$ km.

Figure 3 shows $Q_J, Q_{vis}$, and $Q_{comp}$, and a panel showing their volume integrals together with the magnetic energy, which is the volume integral of $B^2/8 \pi$. The plots of the heating rates suggest that different heating and cooling processes are important at different times during the acceleration process. Mainly for $\bar{t} \lesssim 0.8$, and some intervals in the range $\bar{R} \lesssim 1$, $Q_J \gtrsim Q_{vis}$, but otherwise $Q_{vis}$ tends to significantly dominate $Q_J$. The plots of the volume integrated Joule and viscous heating rates show that Joule heating dominates viscous heating for $\bar{t} \lesssim 0.5$, but not by much.
Overall, compressive heating and cooling dominate the heating effect of $Q_J$ and $Q_{vis}$. $Q_{comp}$ is relatively large and positive for $\bar{t} \lesssim 0.5$, and negative and relatively large in magnitude at later times, suggesting a strong cooling effect due to rarefaction $(\nabla \cdot {\bf V} > 0 \Rightarrow d\rho/dt <0)$. Characteristic mean chromospheric heating rates, assuming emission over a height range $\sim 10^3$ km, are $\sim (4 \times 10^6 - 2 \times 10^7)$ ergs-cm$^{-2}$-s$^{-1}$/$10^3$ km = $0.04 - 0.2$ ergs-cm$^{-3}$-s$^{-1}$ (Withbroe \& Noyes 1977; Vernazza, Avrett \& Loeser 1981; Anderson \& Athay 1989). Then, locally and for tens of seconds, the estimated values of $Q_J, Q_{vis}$, and especially the magnitude of $Q_{comp}$ are comparable to, or far exceed the mean chromospheric heating rate.  

Determining $E_J, W, E_{vis}$, and $E_{comp}$ as in \S 6.1.1 gives $(E_J, W, E_{vis}, E_{comp})= (2.48 \times 10^{20}, 4.23 \times 10^{21}, 2.98 \times 
10^{21}, -8.16 \times 10^{21})$ ergs. About $95 \%$ of the magnetic energy of $E_J+W=4.48 \times 10^{21}$ ergs that is expended in accelerating the plasma, and heating it by resistive dissipation is converted into bulk flow kinetic energy. The value of $E_J + W$ is of the same order of magnitude as the larger values of the magnetic energy $(\sim 10^{21} \; \mbox{ergs})$ in Fig. 3. This is consistent with the fact that the magnetic field provides the energy for the acceleration and heating process. About 98\% of $E_J$ is due to Pedersen current dissipation. Viscous heating exceeds Joule heating by a factor $\sim 12$. The net compressive heating rate is relatively large and negative, indicating a net cooling effect during the acceleration process. 

The energy estimates presented here are crude because they are not constrained by an energy equation. The question of the degree to which these estimates are realistic can be addressed in a meaningful way only by using models that include an energy equation. That equation couples $Q_J, Q_{comp}$, and 
$Q_{vis}$ to one another, and to radiative cooling, and diffusive and convective thermal energy flow. Solving an energy equation self-consistently as part of the model is also the only way to obtain a meaningful estimate of the variation of $T$ in space and time.  

\subsubsection*{6.1.3. Energy and Mass Fluxes into the Corona}

 Ji et al. (2012) report observations of impulsive coronal heating associated with upward mass flows originating in photospheric regions of strong magnetic field in intergranular lanes. The flows from the photosphere into the corona are observed to occur along magnetic loops with diameters $\sim 100$ km, comparable to the reported diameters of type 2 spicules. Some of these flows are observed as RBE's. The observations of Ji et al. are consistent with the suggestion that type 2 spicules are an important source of mass and energy for the corona. Semi-empirical analyses of De Pontieu et al. (2011) and Klimchuk (2012) respectively support and refute the possibility that type 2 spicules make a significant contribution of mass and energy to the corona.

The question of how much energy and mass type 2 spicules contribute to the corona is addressed here by computing the horizontal area and time averaged vertical mass flux, and electromagnetic, convective thermal, and bulk kinetic energy fluxes for the solutions presented here. These fluxes are evaluated at the height above the reference height $(z=0)$ at which the density is $10^{11}$ cm$^{-3}$, defined here as the top of the chromosphere. They are upper limits to the mass and energy fluxes into the corona. 

Let $R_\ast = 100$ km, and $t_\ast = 4 t_0$. The average mass flux over the area $\pi R_\ast^2$, and time interval $0 \leq t \leq t_\ast$ is
\begin{equation}
F_M = \frac{2}{R_\ast^2 t_\ast} \int_0^{t_\ast} dt \int_0^{R_\ast} dx \; \rho V_z x. \label{500}
\end{equation}
The average convective thermal energy flux $F_{TE}=3 k_B T F_M/(2 m_p)$.

The average bulk kinetic energy flux is 
\begin{equation}
F_{KE} = \frac{1}{R_\ast^2 t_\ast} \int_0^{t_\ast} dt \int_0^{R_\ast} dx \; \rho V_z^3 x. \label{101}
\end{equation}

The average electromagnetic energy flux $F_{EM}$ is computed using the $z$ component $S_z = c(E_R B_\theta - E_\theta B_R)/(4 \pi)$ of the Poynting flux. Equations (\ref{300}) and (\ref{100}) are used to compute $E_R$ and $E_\theta$. Then
\begin{equation}
F_{EM} = \frac{2}{R_\ast^2 t_\ast} \int_0^{t_\ast} dt \int_0^{R_\ast} dx \; S_z x. \label{102}
\end{equation} 

Given the estimated total energy flux $F_t= F_{TE}+F_{KE}+F_{EM}$ into the corona due to a single spicule, and assuming the spicules occur with a frequency $f$ such that they balance a coronal energy loss flux $F_c$ over the entire surface area $A_s$ of the Sun, the type 2 spicule occurrence frequency for a given solution is estimated as
\begin{equation}
f=\left(\frac{F_c A_s 100 \; \mbox{s}}{t_\ast F_t \pi R_\ast^2}\right) \left(100 \; \mbox{s}\right)^{-1} \label{200}.
\end{equation} 
Here $f$ is expressed in number per 100 s because this is a characteristic observed spicule lifetime. Choose $F_c = 10^6$ ergs-cm$^{-2}$-s$^{-1}$. 

For the solution in \S 6.1.2, $(F_M, F_{TE}/F_t, F_{KE}/F_t,F_{EM}/F_t, F_t) = (1.33 \times 10^{-7} \; \mbox{g-cm$^{-2}$-s$^{-1}$},14.9 \%,$ $70.3 \%,
14.8 \%, 8.88 \times 10^5 \; \mbox{ergs-cm$^{-2}$-s$^{-1}$})$, and $f= 1.64 \times 10^8 \;(100 \; \mbox{s})^{-1}$. A characteristic solar wind mass flux 
$F_{sw}$ at the base of the corona is $10^{-11}$ g-cm$^{-2}$-s$^{-1}$ (Withbroe \& Noyes 1977). Then $F_M \sim 10^4 F_{sw}, F_{KE}$ dominates
the energy flux, and $F_t$ is comparable to the average coronal energy loss during a time $t_\ast$. The value of $F_M$ implies that if all of $F_t$ contributes to balancing coronal energy loss, then essentially all the spicule mass injected into the corona must return to the chromosphere, but this
mass must be in the corona long enough to transfer all of its energy to coronal plasma. The value of $f$ is $\sim 10$ times larger than that inferred from the semi-empirical estimate of Judge \& Carlsson (2010) that there are $\sim 2 \times 10^7$ type 2 spicules distributed over the surface of the Sun at any given time. 

The value $F_c = 10^6$ ergs-cm$^{-2}$-s$^{-1}$ used above is a characteristic value for the coronal energy loss. The value of this loss varies over the surface of the Sun. Withbroe \& Noyes (1977) give values of $F_c = (3 \times 10^5, 8 \times 10^5, 10^7)$ ergs-cm$^{-2}$-s$^{-1}$ for quiet Sun, coronal hole, and active regions. Using these values in Eq. (\ref{200}) gives values for $f \sim 2.5, 6.6$, and 82 times larger than the value inferred from Judge \& Carlsson (2010). Given observational and model uncertainties, this suggests type 2 spicules may make a significant contribution to coronal energy input in quiet Sun regions, possibly also in coronal holes, but not in active regions. 

\subsection*{6.2. Solution 2}

For this solution $b_{\theta 0}=156.7$ G and $R_0=1.25$ km. Then $B_\theta \leq 25$ G. The increase in $b_{\theta 0}$ by a factor of 2 requires that $R_0$ be reduced by a factor of 4 to keep $|V_R| < V_s$. For the BG state $-0.3 \leq V_z (\mbox{km-s$^{-1}$}) \leq 0$ and $-2.4 \leq 
V_R(\mbox{m-s$^{-1}$}) \leq 0$. Plots for Solution 2 are not included because they are similar in shape to those for Solution 1. The flow is localized within a diameter $\sim 50$ km. The maximum of $V_z$ is 77.5 km-s$^{-1}$. If $\nu=0$, the
maximum of $V_z$ is 397.2 km-s$^{-1}$. The maximum values of $n, Q_J, Q_{vis}$, and $Q_{comp}$ during the main phase of the acceleration are roughly 4, 30-40, 10, and 4-5 times their values for Solution 1.  

Here $(E_J, W, E_{vis}, E_{comp})= (7.38 \times 10^{20}, 1.93 \times 10^{21}, 2.72 \times 10^{21}, -3.26 \times 10^{21})$ ergs.
Then $\sim 72 \%$ of the magnetic energy of $E_J + W =2.67 \times 10^{21}$ ergs that is expended in accelerating the plasma and heating it by resistive dissipation is converted into bulk flow kinetic energy. About 97\% of $E_J$ is due to Pedersen current dissipation. The total thermal energy generated by Joule and viscous dissipation is $\sim 6 \%$ greater than the net cooling due to rarefaction.

The estimated mass and energy fluxes into the corona are $(F_M, F_{TE}/F_t, F_{KE}/F_t,F_{EM}/F_t,$ $F_t) = (4.88 \times 10^{-8} \; 
\mbox{g-cm$^{-2}$-s$^{-1}$},14.8 \%,65.6 \%,$ $19.6 \%, 3.27 \times 10^5 \; \mbox{ergs-cm$^{-2}$-s$^{-1}$})$, and $f= 4.45 \times 10^8 \;(100 \; 
\mbox{s})^{-1}$ for $F_c = 10^6$ ergs-cm$^{-2}$-s$^{-1}$. The energy injected into the corona is $\sim 2.7$ times less than for Solution 1, and the required value of $f$ is correspondingly $\sim  2.7$ times larger. The larger contribution of Solution 1 to the coronal energy input is due to the larger diameter of the region where the acceleration and heating are concentrated. An important unknown quantity is the occurrence frequency of spicules as a function of their diameter and energy content.

\section*{7. Conclusions}

The HI viscous force can have a strong braking effect on type 2 spicule acceleration if it is driven by Lorentz forces associated
with current densities localized within diameters $\sim 5 - 20$ km, corresponding to $4 R_0$ for the values of $R_0$ used here. If the actual characteristic radii within which the current density is localized are much larger, then the effect of viscous braking is expected to be
significantly smaller. This follows from a dimensionless form of the momentum Eq. (\ref{3}), which shows that the Lorentz 
force is $\propto R_0$, while the viscous force is $\propto R_e^{-1}$, where $R_e = \rho_c R_0^2/t_0 \nu$ is a Reynold's number, and $\rho_c$ is a characteristic density. Then if all other input parameters are held constant, the ratio of the Lorentz force to the viscous force is $\propto R_0^3$.

The maximum vertical speeds of $\sim 66$ and 77 km-s$^{-1}$ for the solutions presented here are at the lower end of type 2 spicule speeds. Significantly higher speeds might be possible in models not restricted to using values of $R_0$ sufficiently small so the effects of $V_R$ in the momentum equation can be neglected. 

Joule and compressive heating dominate during the early stage of acceleration. Viscous heating dominates at later times when velocity gradients are largest, but the cooling rate due to rarefaction, corresponding to $Q_{comp} <0$, more than cancels this heating rate. The total thermal energy generated by viscous dissipation during the acceleration process is an order of magnitude larger than that due to Joule dissipation, but the total cooling due to rarefaction can be comparable to or larger in magnitude than this energy. These are crude estimates since the model does not include an energy equation.

If all the upward spicule energy flux in the upper chromosphere contributes to coronal energy input, the energy injected by type 2 spicules into the corona may provide a significant fraction of the coronal energy input in quiet regions, possibly also in coronal holes, but not in active regions.

Magnetic flux emergence through the photosphere in inter-granular lanes is a likely source of non-force free current systems that accelerate type 2 spicules through the associated Lorentz force. This conclusion is based on the following complementary sets of observations, and on the MHD simulations cited in \S 2.2. (1) Type 2 spicules are localized on scales $\lesssim 10^2$ km perpendicular to their direction, consistent with inter-granular lane spatial scales. (2) Magnetic flux continually emerges through the photosphere in inter-granular lanes, with quiet Sun field strengths up to several hundred Gauss in internetwork, 
$\sim 10^3$ G in network, and with the emergence first appearing as a region of mixed polarity, nearly horizontal flux (Lites et al. 1996; Orozco 
Su\'{a}rez et al. 2007; Lites et al. 2008; Lites 2009). The emerging magnetic structures become more vertical and quasi-cylindrical as they rise, similar to the geometry of type 2 spicules. (3) On the orders of magnitude larger scales of emerging sunspots, pores, and active regions, ${\bf B}$ emerges carrying currents generated below the photosphere (Leka et al. 1996; Lites et al. 1998; Burnette et al. 2004; Sun et al. 2012). Flaring phenomena in these regions include acceleration of plasma up to speeds $\sim 10-10^2$ times greater than those of type 2 spicules. The field emerges as mixed polarity, nearly horizontal flux ropes with bipolar geometries, and field strengths $\sim 200-600$ G. The topology becomes more complex as the field rises into the chromosphere. Lites et al. (1998) emphasize that sections of these flux ropes can attain kilo-Gauss strength after they become nearly vertical as they flow out of the flux emergence region and rise into the chromosphere. This vertical, quasi-cylindrical geometry is similar to type 2 spicule geometry on smaller scales. By analogy, the observations of flux emergence with plasma acceleration on these relatively larger, better resolved scales support the proposition that emerging non-potential inter-granular flux is a source of current systems that drive type 2 spicules. At present there do not appear to be any other type 2 spicule acceleration mechanisms that are as strongly supported by observations. 

The chromosphere is a highly resistive medium with respect to the Pedersen current density ${\bf J}_P$ ($\sim {\bf J}_\perp$ in the chromosphere) due to the combination of weak ionization and strong magnetization, which distinguishes the chromosphere from the weakly ionized, weakly magnetized photosphere, and the strongly ionized, strongly magnetized corona (Goodman 2000, 2004). Here ``highly resistive'' means the Pedersen resistivity $\eta_P 
(\sim \eta \Gamma)$ is orders of magnitude larger than $\eta$. This raises the following questions: (1) If emerging magnetic flux is considered as a driver of type 2 spicules, are the associated currents strong enough to cause type 2 spicule acceleration? (2) If the answer is yes, is the 
${\bf J}_\perp$ associated with emerging flux sufficiently dissipated by Pedersen current dissipation below the height at which spicules form so that the Lorentz force cannot be a major component of the force that accelerates type 2 spicules? At present these questions cannot be answered for the following reasons: (1) The formation heights of type 2 spicules are not known, other than that they form somewhere in the chromosphere. (2) An approximate, meaningful answer to these questions can be obtained from simulations of instances of the flux emergence process that include the essential physics, and use spatial and temporal resolutions sufficient to accurately compute ${\bf J}_\perp$, and the associated Lorentz force and $Q_P$. Such simulations do not yet exist. The 3D simulations of flux emergence by Arber et al. (2007) are a step towards developing such a simulation, and suggest that as flux rises into the chromosphere, there is strong dissipation of ${\bf J}_P$, consistent with general results in Goodman (2000, 2004), and the expectation that the atmosphere becomes increasingly force free with increasing height. 

\begin{acknowledgments}
The author acknowledges support from grants ATM-0650443 and ATM-0848040 from the Solar-Terrestrial Research Program of the National Science Foundation to the West Virginia High Technology Consortium Foundation. This research made use of NASA's Astrophysics Data System (ADS). The author thanks both
referees for thorough and constructive reviews. 
\end{acknowledgments}

\newpage
\appendix

\newpage
\begin{figure}
\figurenum{1}
\epsscale{1}
\plotone{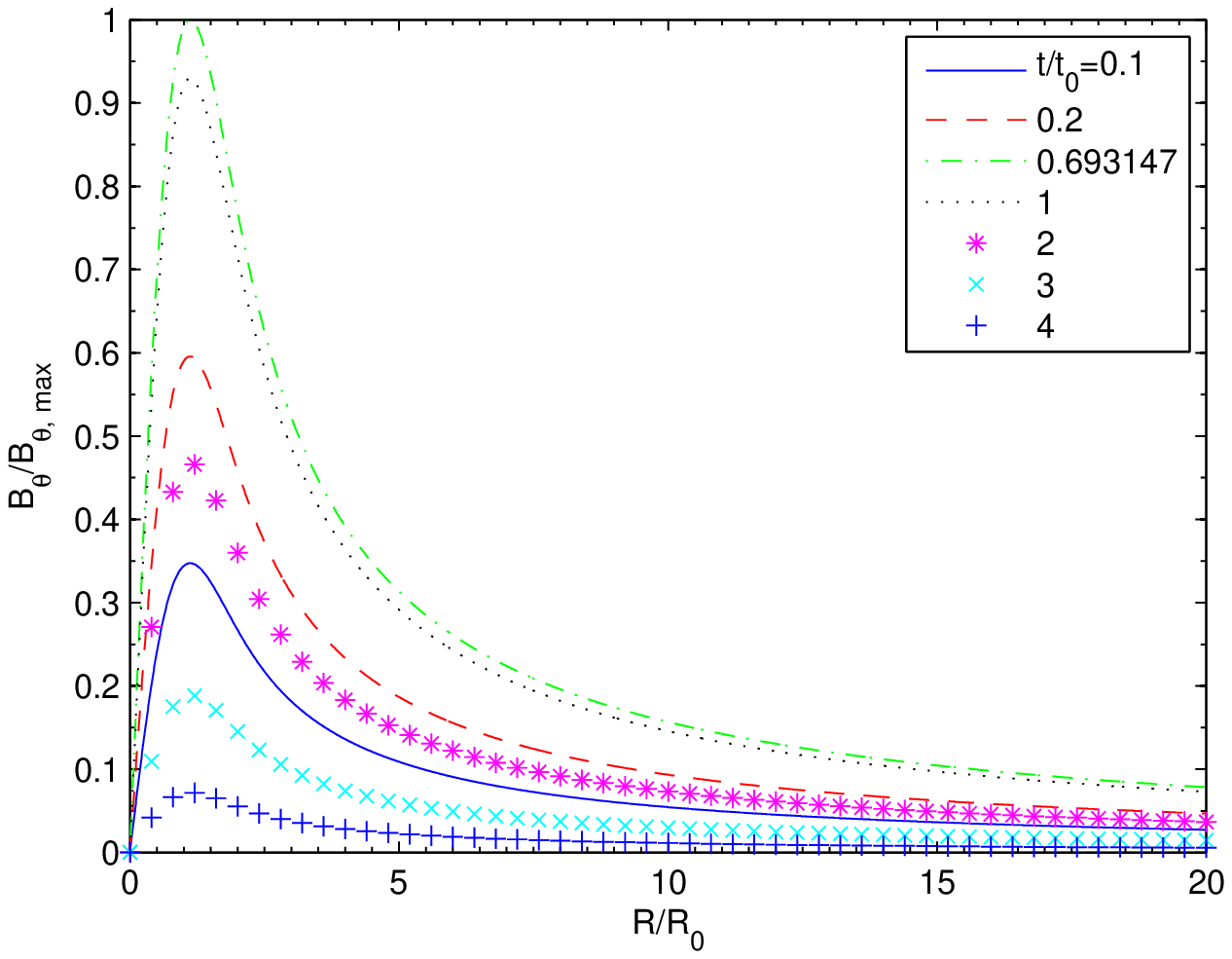}
\caption{Normalized azimuthal magnetic field for all solutions. For Solution 1: $R_0=5$ km, $t_0=33.3$ s, and $B_{\theta, max}=12.5$ G. For Solution 2: 
$R_0=1.25$ km, $t_0=33.3$ s, and $B_{\theta, max}=25$ G.}
\end{figure}

\begin{figure}
\figurenum{2}
\epsscale{1.32}
\plottwo{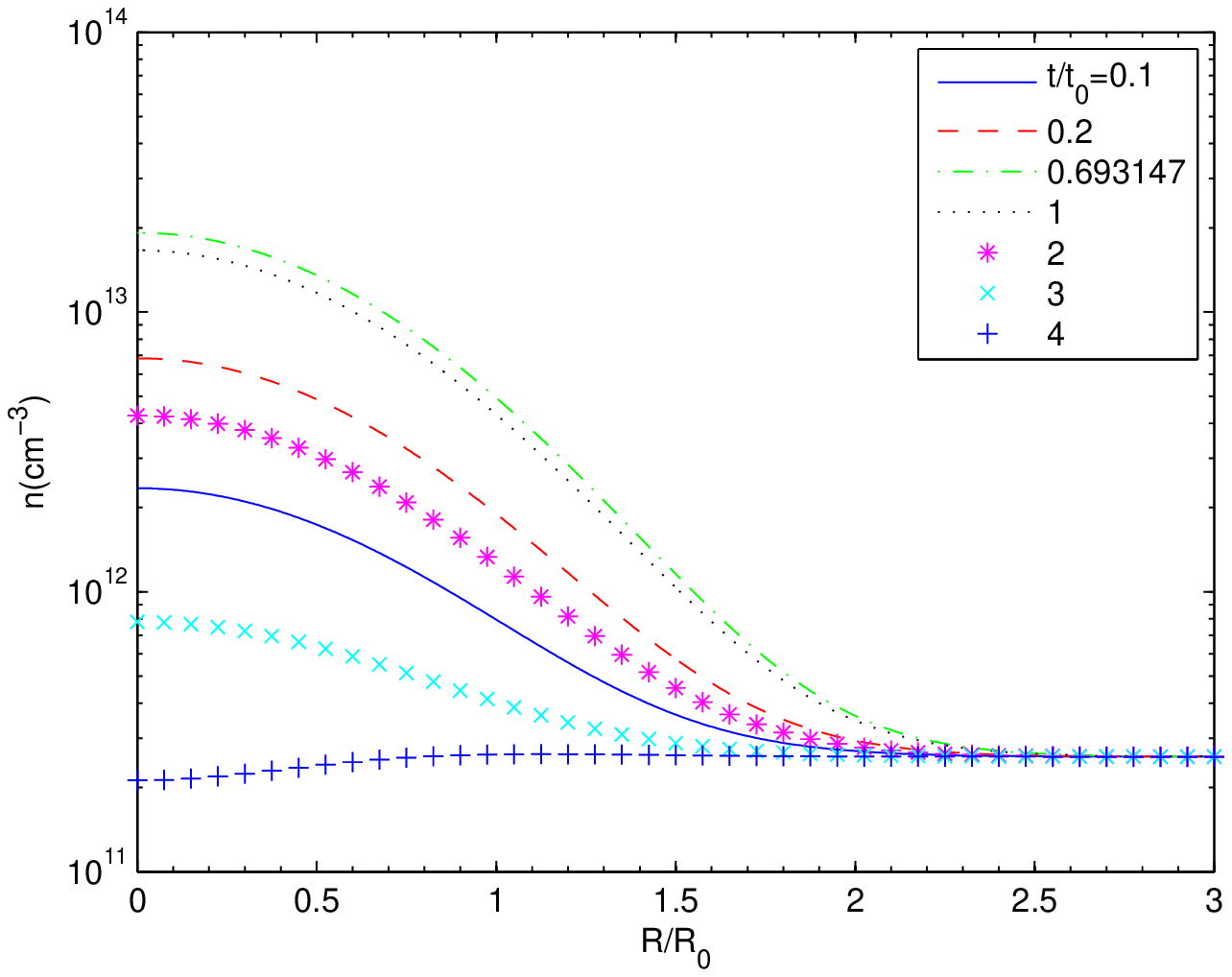}{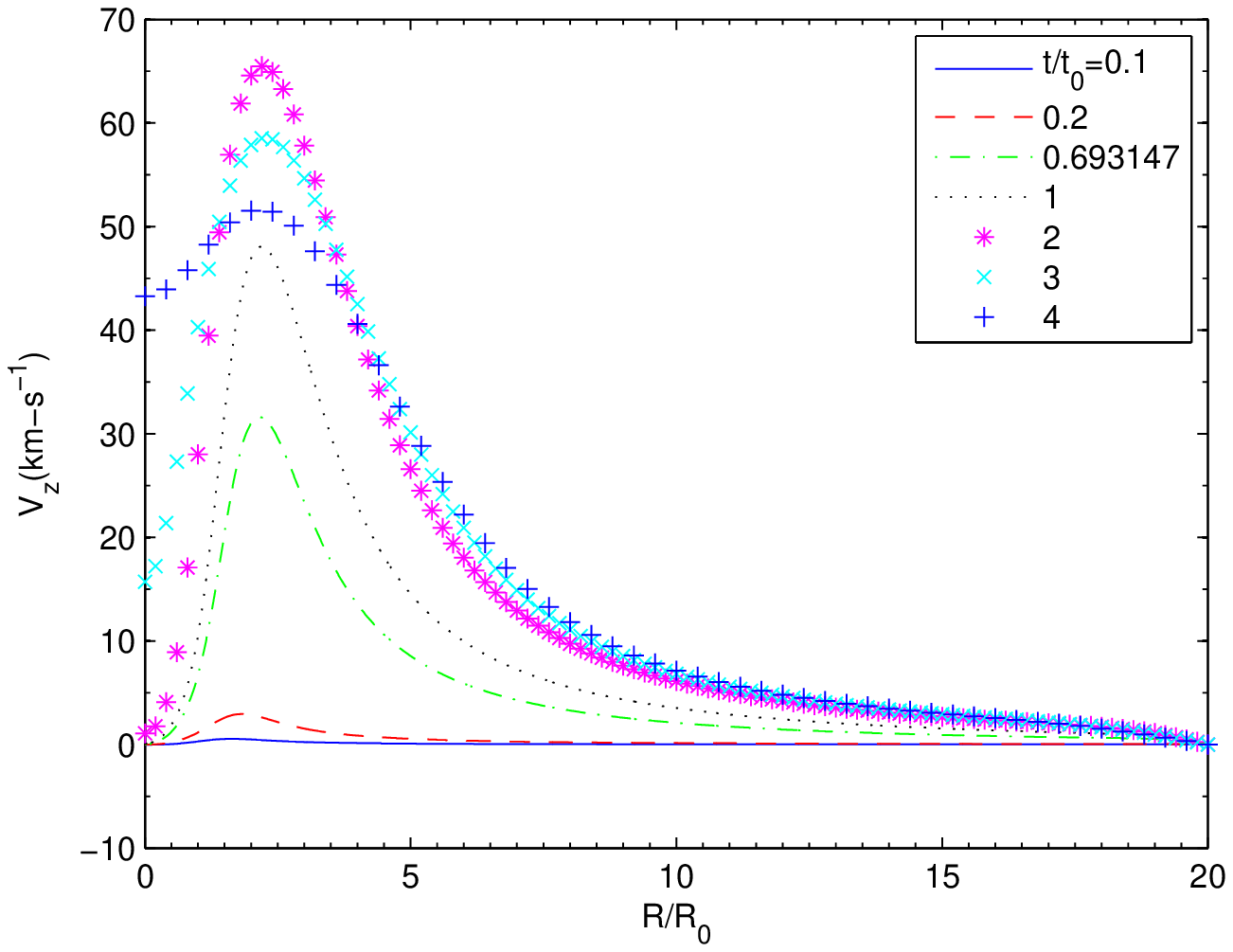}
\epsscale{0.6}
\plotone{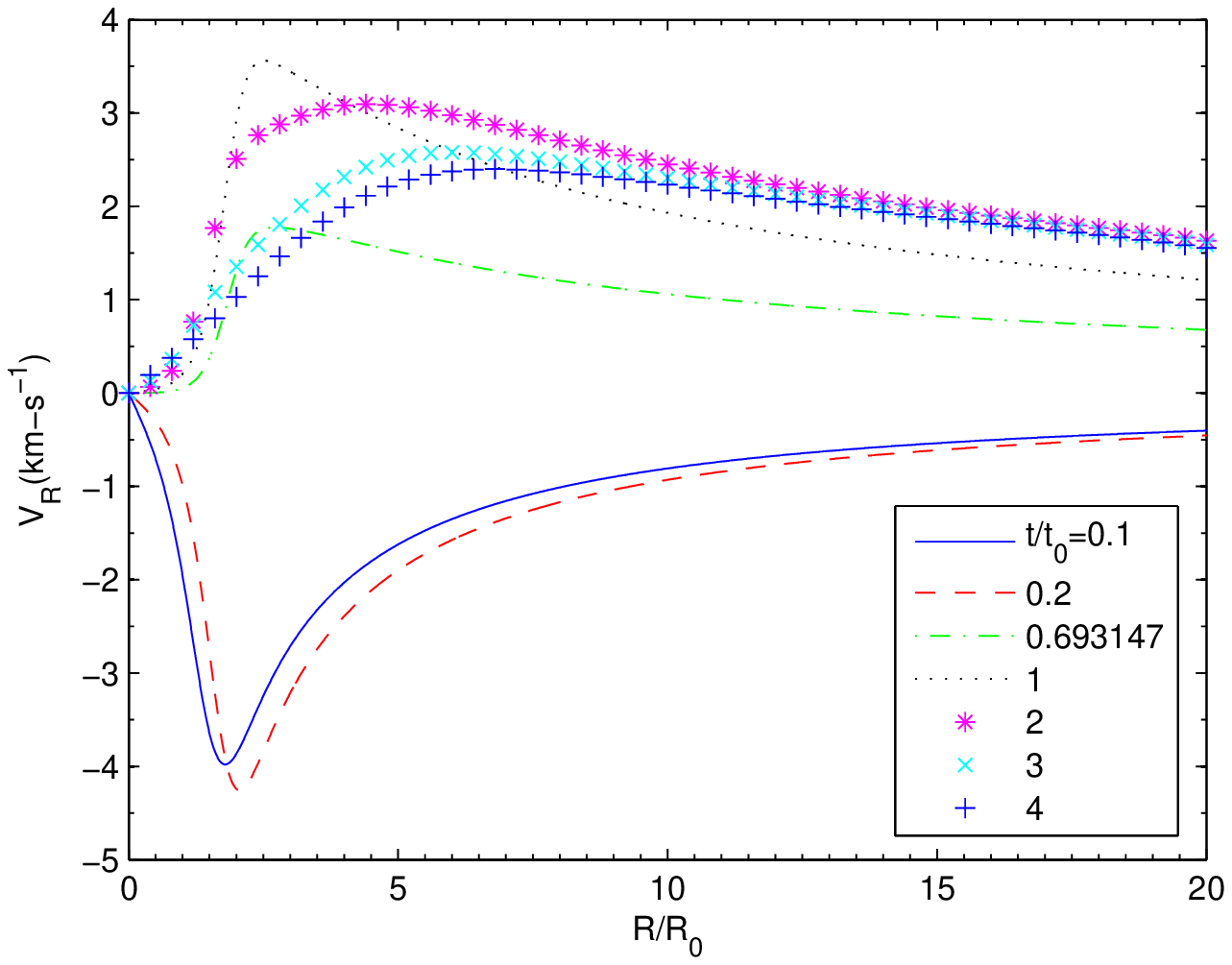}
\caption{Total number density (top), and vertical (middle) and radial (bottom) velocities for Solution 1.}
\end{figure}

\begin{figure}
\epsscale{1.3}
\plottwo{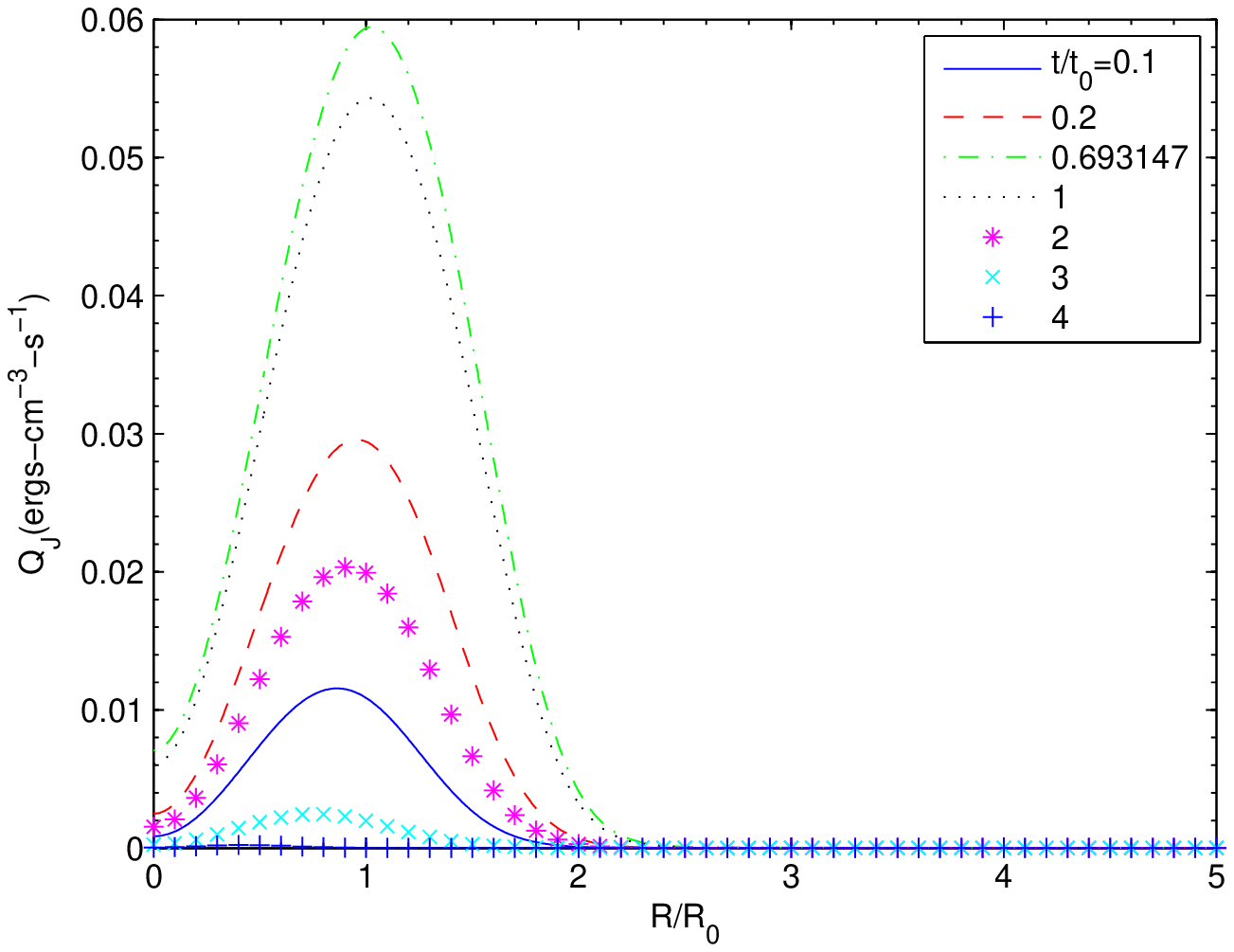}{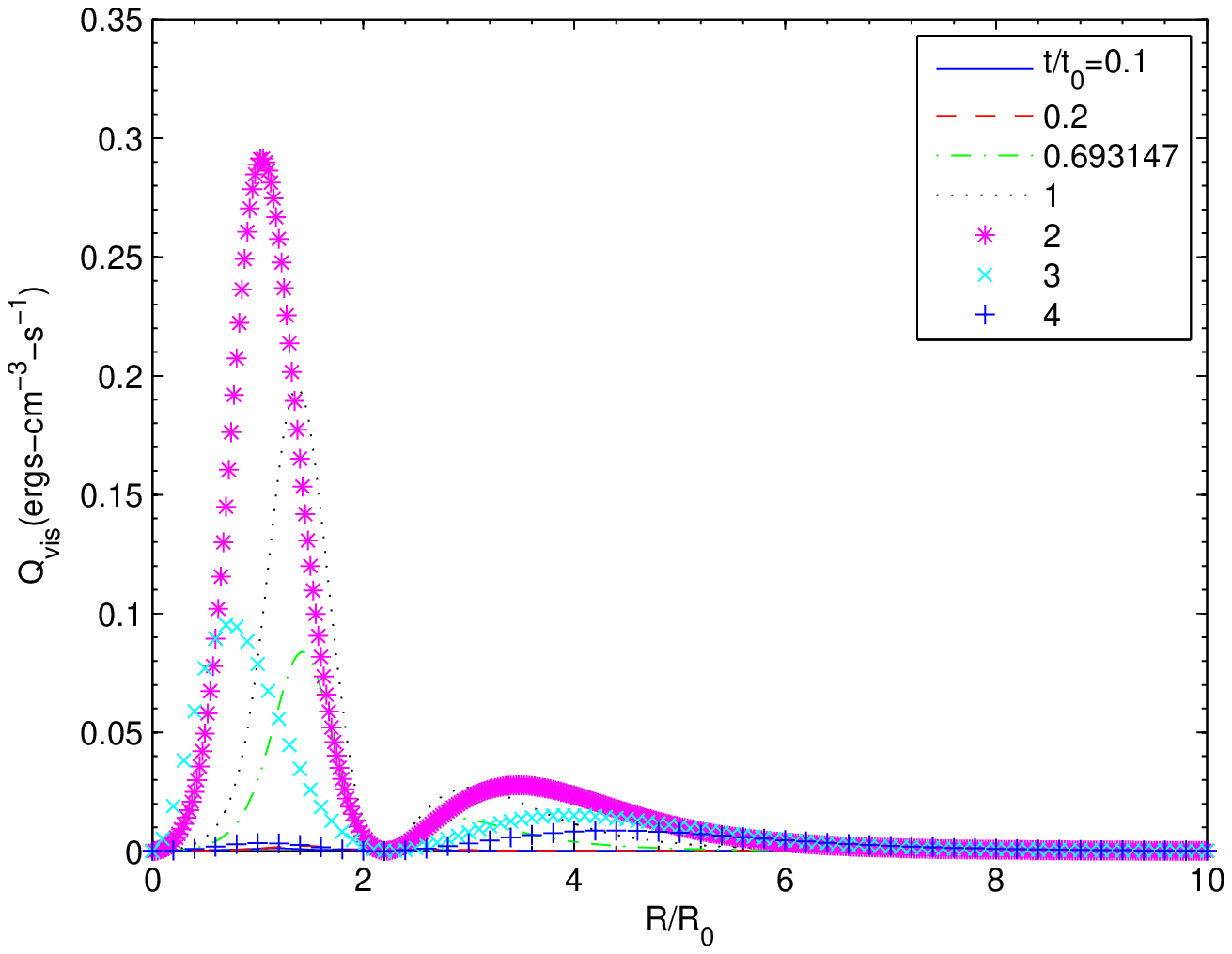}
\figurenum{3}
\plottwo{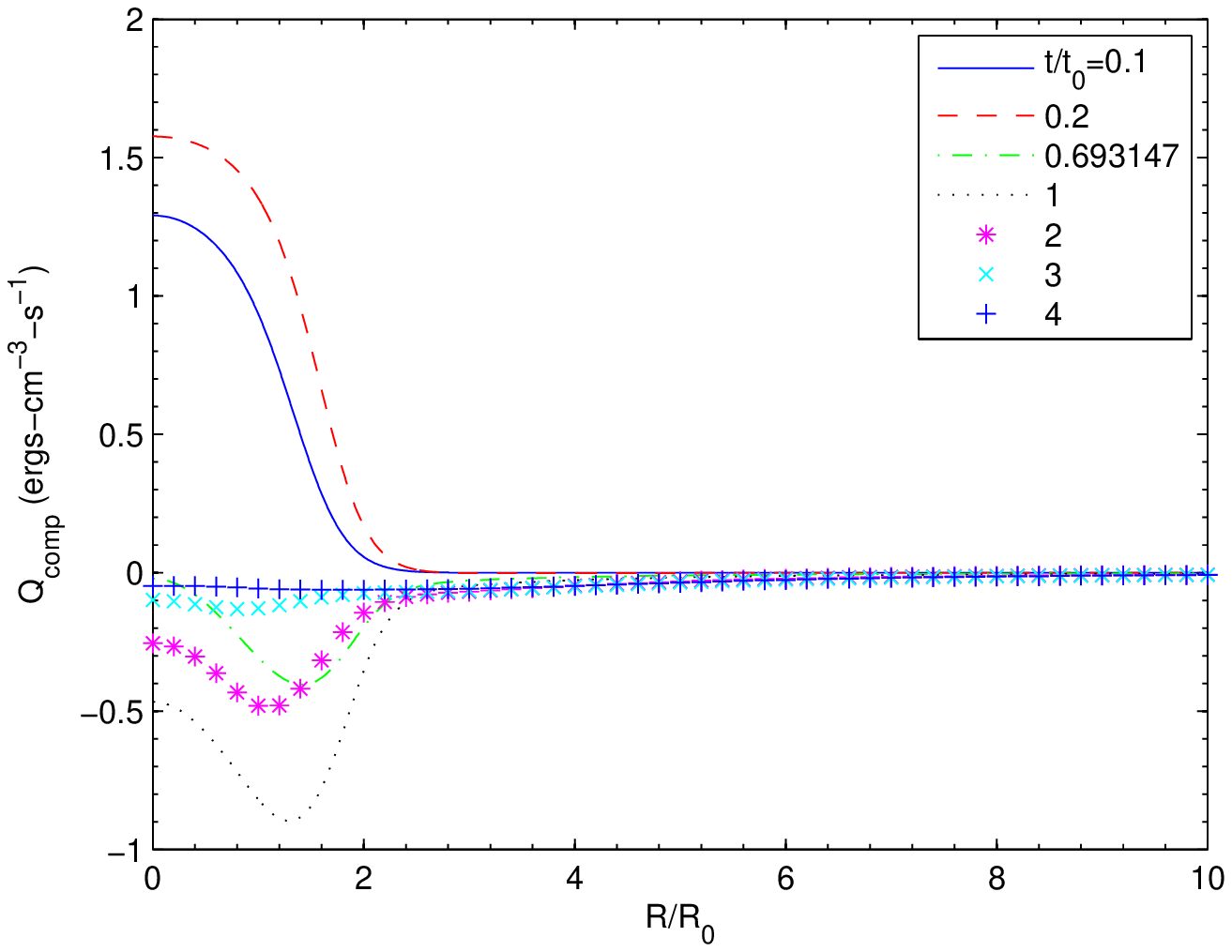}{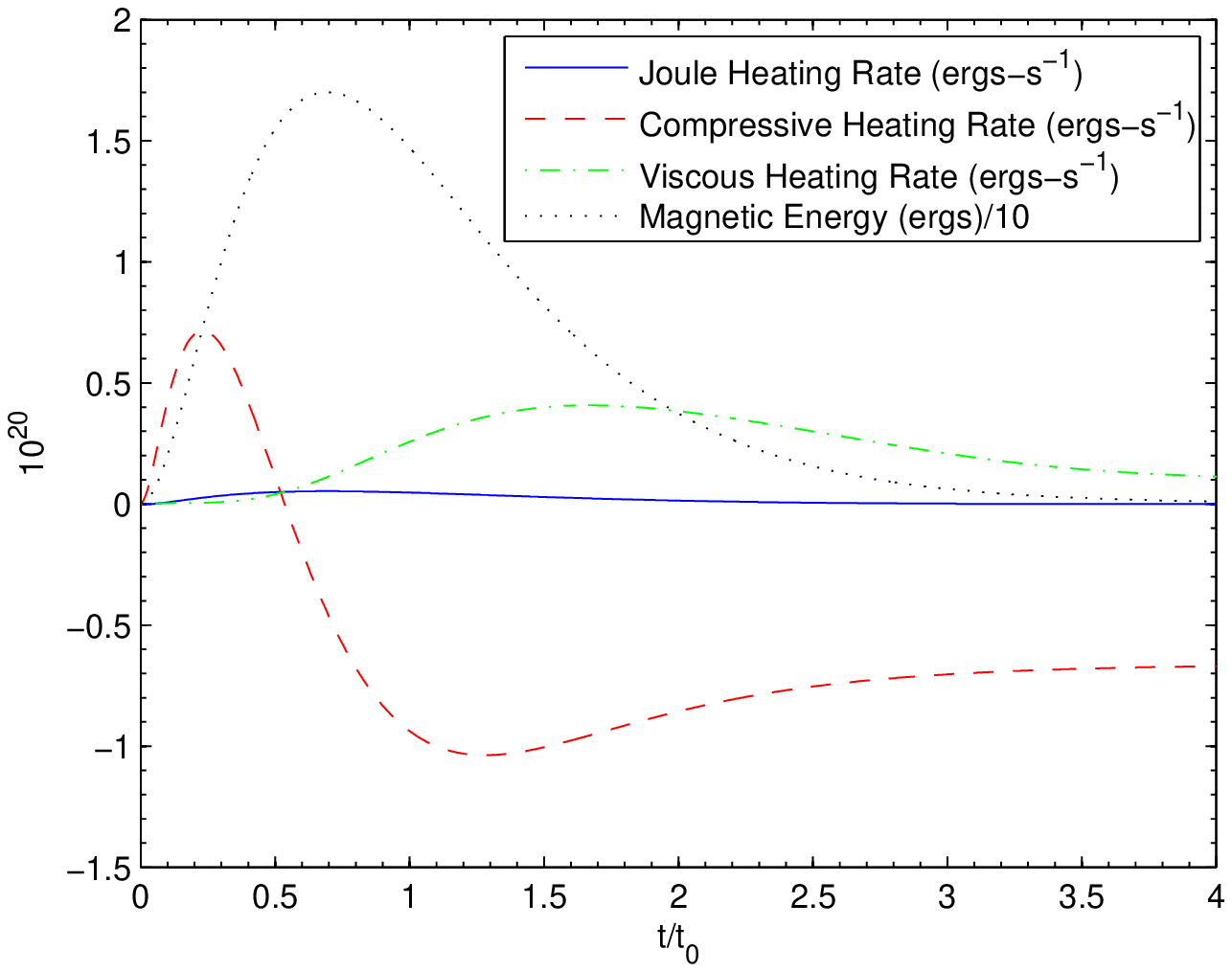}
\caption{Upper left/right: Joule/viscous heating rate per unit volume for Solution 1. Lower left/right: Compressive heating rate per unit 
volume/total magnetic energy for Solution 1.}
\end{figure}

\end{document}